
D\headline={\ifnum\pageno=1\firstheadline\else
\ifodd\pageno\rightheadline \else\leftheadline\fi\fi}
\def\firstheadline{\hfil}
\def\rightheadline{\hfil}
\def\leftheadline{\hfil}
\footline={\ifnum\pageno=1\firstfootline\else\otherfootline\f
i}
\def\firstfootline{\rm\hss\folio\hss}
\def\otherfootline{\hfil}
\font\tenbf=cmbx10
\font\tenrm=cmr10
\font\tenit=cmti10
\font\elevenbf=cmbx10 scaled\magstep 1
\font\elevenrm=cmr10 scaled\magstep 1
\font\elevenit=cmti10 scaled\magstep 1

\font\ninerm=cmr9

\nopagenumbers
\line{\hfil }
\vglue 1cm
\hsize=6.0truein
\vsize=8.5truein
\parindent=3pc
\baselineskip=10pt
\noindent{GR-QC 9504002.}

\noindent{\it ``Quantum Coherence and Reality'',
Proceedings of the Conference on Fundamental Aspects of
Quantum
Theory, Columbia SC, Dec. 1992, to celebrate Yakir Aharonov's
60th
Birthday, edited by J. S. Anandan and J. L. Safko (World
Scientific
1994).}
\vglue 1.0cm
\centerline{\tenbf TOPOLOGICAL PHASES AND THEIR
DUALITY IN}
\vglue 12pt
\centerline{\tenbf ELECTROMAGNETIC AND GRAVITATIONAL
FIELDS}
\vglue 5pt
\vglue 1.0cm
\centerline{\tenrm J. ANANDAN }
\baselineskip=13pt
\centerline{\tenit Department of Physics and Astronomy,
University of South Carolina}
\baselineskip=12pt
\centerline{\tenit  Columbia, SC 29208, USA}
\vglue 0.8cm
\centerline{\tenrm ABSTRACT}
\vglue 0.3cm
{\rightskip=3pc
 \leftskip=3pc
 \tenrm\baselineskip=12pt
 \noindent
The duality found by Aharonov and Casher for topological
phases in
the
electromagnetic field is generalized to an arbitrary linear
interaction.
This
provides a heuristic principle for obtaining a new solution of
the field
equations from a known solution. This is applied to the
general
relativistic
Sagnac phase shift due to
the gravitational field in the interference of mass or energy
around a
line
source that has angular momentum and the dual phase shift
in the
interference of a spin around a line mass. These topological
phases
are
treated both in the
linearized limit of general relativity and the exact solutions
for which the gravitational sources are cosmic strings
containing
torsion
and curvature, which do
not have a Newtonian limit.

\vglue 0.8cm }
\line{\elevenbf 0. Introduction \hfil}
\bigskip
\elevenrm

As is well known, some of Yakir Aharonov's most famous
contributions concern topological phases due to the
electromagnetic
field. It is therefore fitting on this occasion of his sixtieth
birthday
for me to present him with some observations concerning
these
phases, which generalize naturally to the gravitational field.
In
particular, I shall examine the duality between the Aharonov-
Bohm
(AB) phase [1]  and the phase shift in the interference of a
magnetic
moment in an electric field [2] which was found by Aharonov
and
Casher (AC) [3]. I shall show, by means of the linearized limit
and an
exact
solution of the
gravitational field equations, that both these phases have
gravitational analogs and they satisfy this duality.

In section 1, I shall briefly summarize the phase shifts in the
interference of a charge and a magnetic dipole (at low
energies) due
to the electromagnetic field. These phase shifts reveal,
respectively,
$U(1)$ and $SU(2)$ gauge field aspects of the electromagnetic
field.
But these two aspects are not independent: The $SU(2)$
connection
which gives the dipole phase shift depends on the electric and
magnetic fields
and as
such are derived from the electromagnetic connection
that gives the $U(1)$ AB phase shift. It is nevertheless
amusing to
see a charged particle with a magnetic moment, such as an
electron,
interacting with an
electromagnetic field as if it is a $U(1)\times SU(2)$ gauge
field. Two
topological phase shifts due to electric and manetic fields
corresponding to two $U(1)$ subgroups of
$SU(2)$ will be reviewed.
The duality between one of these and the AB phase, found by
AC,
will be generalized to an arbitrary interaction in section 2. I
shall
formulate a duality principle which states that any two dual
phases
are equal under certain conditions.

The gravitational phase shifts, obtained in section 3, are
special cases
of the phase shifts obtained previously [4, 5] due to the
coupling of
the mass and spin to the gravitational field. The key to the
analogy
with the electromagnetic phase shifts is that the mass or
energy
plays the role of the electric charge and spin the role of the
magnetic
dipole in the electromagnetic field. The gravitational phase
shifts are
the same as due to the phase shifts of a Poincare gauge field.
The
translational and Lorentz
aspects of the Poincare group are respectively analogous to
the
$U(1)$ and $SU(2)$ aspects of the electromagnetic field,
mentioned
above.

If gravity contains torsion, as will be assumed here, the
connection,
which gives the phase shift of the spin, is independent of the
metric
or the vierbein, which gives the phase shift due to the mass
or
energy-momentum.
Therefore, these two aspects are then complementary, unlike
in the
electromagnetic case in which the $SU(2)$ connection depends
on the
$U(1)$ connection as mentioned earlier. The electromagnetic
field
and its sources of
course must satisfy the Maxwell's equations. It is well known
that
the solenoid which produces the AB phase shift is a solution.
Similarly,
the gravitational field and its sources must satisfy Einstein's
field
equations or a suitable generalization of it to include torsion.
Fortunately, an exact solution corresponding to a spinning
cosmic
string with angular momentum and mass, which is the analog
of the
solenoid
with a coaxial line charge
in the electromagnetic case, can be obtained everywhere
including
the interior of the string.

There is a topological general relativistic
Sagnac phase [4] which depends on the energy of a particle
outside the string and the flux of torsion inside the string
produced
by
its spin. This is analogous to the AB phase. There is another
topological phase which depends on the spin of the particle
and the
flux of curvature inside the string, produced by its
mass. This is the dual of the former phase. I
shall show that this pair of dual phases satisfy the duality
principle
formulated in section 2.

\vglue 0.6cm
\line{\elevenbf 1. Topological and Geometrical Phases due to
the
Electomagnetic Field \hfil}
\vglue 0.4cm

For simplicity, consider the non relativistic Hamiltonian of a
charged
particle in an electromagnetic field
$$H = {1\over 2m}({\bf p}-e{\bf A})^2 + eA_o , \eqno(1.1)$$
where $e$ and $m$ are the charge and mass of the particle
and
$A_\mu = (A_0, A_i)= (A_0,-A^i)$ is the electromagnetic
potential
representing the
$U(1)$ connection due to this
field\footnote{*}{\ninerm\baselineskip=11pt Units in which
the
velocity of light
$c=1$ will be used throughout.}. As is well known, (1.1)
predicts
the Aharonov-
Bohm (AB) effect [1]. This is the electromagnetic phase
difference
between two interfering coherent beams which are entirely in
a
multiply connected region in which the field strength
$F_{\mu \nu}$
is zero. The
phase factor that determines the electromagnetic shift in the
interfering fringes is
$$\Phi_C =  \exp(- {ie\over \hbar}\oint_{C} A_\mu
dx^\mu), \eqno(1.2)$$
where the closed curve $C$ passes through
the two interfering wave functions, and encloses a region in
which
the field
strength is non zero. Consequently, this phase shift is
constant as
the beams are varied in the outside region in which the field
strength is zero, which makes this effect topological.
Conversely, the
experimental observation of this phase shift may be used to
infer
that the electromagnetic field is a $U(1)$ gauge field [6,7].

Consider now the interaction of a neutral magnetic dipole,
such as a
neutron, with the electromagnetic field which at low energies
is
described by the Hamiltonian [2,3,8,9]
$$H = {1\over 2m}({\bf p}- \gamma {\bf A}^k S_k)^2 +
\gamma
{A_o}^kS_k , \eqno(1.3)$$
where ${A_\mu}^k = ({A_0}^k,{A_i}^k)=(-B^k, \epsilon_{ijk}
E^j)$ in
terms of the electric field ${\bf E}$ and the magnetic field
${\bf B}$,
$S_k, k=1,2,3$ are the spin components which generate the
$SU(2)$
spin group. For a spin $1\over 2$ particle, the magnetic
moment
$\mu ={\gamma
\hbar\over2}$. This interaction is like that of an isospin with
an
$SU(2)$ Yang-Mills field [8,9,10].

The phase shift due to both electric and magnetic fields, in the
interference of a neutral dipole such as a neutron, was
obtained by
means of an explicit plane wave solution [2]. This result, of
course,
applies also to the more general situation when the
interfering wave
functions are locally approximate plane waves so that the
WKB
approximation is valid.
Hence,
the phase shift is determined by
the non abelian phase factor
$$\Phi_C =  P\exp(- {i\gamma\over \hbar}\oint_C
{A_\mu}
^kS_k dx^\mu),\eqno(1.4)$$
where P denotes path ordering, and
$C$ is a closed curve consisting of unperturbed trajectory [8].
Hence,
$\Phi_C$ is an element of $SU(2)$, and this phase shift is like
the phase shift due to an $SU(2)$ gauge field [5,6,11].

The special case when the two waves interfere around a line
charge
was considered by AC [3]. In this case, ${A_0}^k =
0$ and the electric field $E^j$ and therefore ${A_i}^k
=\epsilon_{ijk}
E^j$ fall off inversely as the distance from the line charge. It
follows
immediately from (1.4) that, if the spin is polarized parallel to
the
line charge, then this phase shift is topological in the sense
that it
does not change when the curve $C$ surrounding the line
charge is
deformed. However, the Yang-Mills field strength
${F_{\mu\nu}^k}$
of ${A_\mu}^k$ is non vanishing outside the line charge,
which
makes
this effect fundamentally different from the AB effect in
which the
electromagnetic field strength $F_{\mu\nu}=0$ along the
beams. But
if the line charge is in the 3-direction then
${F_{\mu\nu}^3}=0$. That
is the field strength corresponding to the $U(1)$ subgroup of
the spin
$SU(2)$ generated by $S_3$ is zero. So, for this subgroup, this
phase
shift is like the AB effect.

Another topological phase shift experienced by the dipole is
the
following: The wave packet of a neutron or an atom is split
into two
coherent wave
packets, one of which enters a cylindrical solenoid. The
homogeneous
magnetic field of the solenoid is then turned on and is then
turned
off before
the wave packet leaves the solenoid. Then there is a phase
shift even
though there is no force acting on the neutron. This phase
shift,
which is easily obtained from (1.4), is due to ${A_0}^k=-B^k$.
Hence,
this phase shift is due to the scalar potential of the gauge field
of the
$U(1)$ subgroup of the spin $SU(2)$ group generated by the
component of
${\bf S}$ in
the direction of ${\bf B}$. At the suggestion of Zeilinger [12]
and the
author [8] this experiment was performed for neutrons [13].

The general case of the phase shifts for a particle that has
charge and
magnetic moment
interacting with an electromagnetic field was studied before
[8,9]. I
shall restrict here, for simplicity, the special case of the
particle being
a Dirac electron with ``g-factor'' being two. Its Hamiltonian, at
low
energies in the inertial frame of the laboratory, is
$$H = {1\over 2m}({\bf p}- e{\bf A}-{1\over 2}\gamma '{\bf
A}^kS_k)^2 + eA_o + \gamma '{A_o}^kS_k , \eqno(1.5)$$
This Hamiltonian is like as if the electron is interacting with a
$SU(2)\times U(1)$ gauge field represented by the gauge
potentials
${\bf A}^k$ and ${\bf A}$. Note, however, the factor $1\over
2$ in
front of ${\bf A}^k$ compared to (1.3). This is due to the
Thomas
precession undergone by the electron when it accelerates in
the
electric field [2,9].

\vglue 0.6cm
\line{\elevenbf 2. The Duality of AC and its Generalization
\hfil}
\vglue 0.4cm

The major new contribution of AC, which is not contained in
any
earlier work, is the recognition that the phase shift due to the
line
charge is ``dual'' to the AB effect due to a solenoid. I shall now
give a
precise statement of this duality which would be general
enough to
apply to other interactions as well.

Suppose that an infinite uniform solenoid is situated along the
z-axis
of a Cartesian coordinate system. A charge of strength $e$ is
taken
slowly around the solenoid along a circle in
the $xy$-plane with its center at the solenoid, which is
assumed to
have negligible cross-section. The solenoid may be
regarded as a magnetized medium with a constant magnetic
moment
per unit length equal to ${\bf M}$, say, which is parallel to
the $z$-
axis. The AB phase shift acquired by the charge is
$$\Delta\phi = {e\over\hbar}\int_\Sigma {\bf B}\cdot d{\bf
\Sigma}={eM\over\hbar} ,
\eqno(2.1)$$
where $\Sigma$ is a cross-section of the solenoid, $\bf B$ is
the
magnetic field inside the solenoid, and $M=|{\bf M}|$.

Now, divide the solenoid into slices each of height $\delta
\ell$
bounded by cross-sections that are parallel to the $xy$-plane.
The
magnetic moment of each slice is
$${\bf \mu}={\bf M}\delta \ell . \eqno(2.2)$$
The linearity of Maxwell's equations imply that the phase
shift is {\it
linear}
in the sense that
(2.1) is the sum of the phase shifts due to the influence of
each slice
of the solenoid on the charge. Consider a slice whose center is
at
$z=Z$. Then taking the charge around the circle mentioned
above is
equivalent to keeping the charge fixed at $z=-Z$ and taking
the slice
of magnetic moment $\mu=|{\bf \mu}|$ around the same
circle in the
$xy$- plane.  The
phase shifts acquired in both processes are the same. This has
been
shown using space-time translation and Galilei invariance
of the Lagrangian, for the special case of charge-dipole
interaction
[3] and using Lorent invariance for the general case of an
arbitrary
interaction [9]. Now do
this for each {\it pairwise} interaction between the charge $e$
and
each slice with magnetic moment $\bf \mu$. Then, as we
account for
all
slices from $z= +\infty$ to $z=-\infty$, in the new situation,
which
will be called the dual of the original situation, there are
charges
from $z= -\infty$ to $z=+\infty$ along the $z$-axis, and the
magnetic
moment $\bf \mu$ circles around this line charge. Each
charge $e$ is
contained in an interval of height $\delta \ell$, and may be
assumed
to be spread uniformly in that interval. Therefore,
$$e=\lambda\delta\ell ,\eqno(2.3)$$
where $\lambda$ is the charge per unit length. It follows that
the
magnetic
moment which circles around this line charge, with its
direction
parallel to the $z$-axis, acquires a phase shift equal to
$\Delta\phi$ given by (2.1). From (2.2) and (2.3),
$${e\over \lambda} ={d\over M}.\eqno(2.4)$$
for these two dual situations.
Using (2.2), (2.1)
may be rewritten as
$$\Delta\phi = {\lambda \mu\over \hbar}. \eqno(2.5)$$
This phase shift may also be independently derived using
(1.4) and
the electric field of a line charge obtained by solving
Maxwell's
equations.

The above argument may be generalized to the case when the
charge
goes around an arbitrary closed curve ${\bf r}(t)$ which may
or may
not enclose the solenoid. Then relative to one of the above
mentioned
slices
at say ${\bf Z}=(0,0,Z)$
this curve is ${\bf r}(t)-{\bf Z}$. Therefore, in the dual
situation the
slice with magnetic moment $\mu$ moves around the closed
curve
$-
{\bf r}(t)+{\bf Z}$ relative to the charge. So, if the charge is
placed at
$-{\bf
Z}$, the slice
goes around the closed curve $-{\bf r}(t)$. By doing this for
each
pairwise
interaction between the charge
and the fixed magnetic moment of each of the slices into
which the
solenoid is divided, I obtain the dual situation in which a
magnetic
dipole of strength $\mu$, and direction parallel to the $z$-
axis,
moves
around the closed curve $-{\bf r}(t)$ with a line charge,
whose
charge per unit length is $\lambda$, along the $z$-axis. Also,
since
this
interaction is invariant under parity, the same phase shift is
obtained
for the situation obtained by parity transforming about the
origin.
This corresponds to the magnetic moment moving around the
original curve ${\bf r}(t)$ traveled by the charge when in the
presence of the solenoid.

Now the statement that the AB phase shift is topological may
be
expressed as follows: If the curve ${\bf r}(t)$ goes around the
solenoid $n$ times, $n=0,1,2,3,...$, then the phase shift
acquired by
the charge going around this curve is $n\Delta\phi$,
independent of
the shape of this curve. (Topology has to do with integers. So,
a
topological phase shift should, strictly speaking, be expressed
in
terms of integers.) Then the curve $-{\bf r}(t)$, which is the
parity
transform of the original curve, goes around the line charge
$n$
times in the dual situation also. Therefore, the phase shift
acquired in
the dual situation is $n\Delta\phi$, independent of the shape
of this
curve. Hence, the latter phase shift is also topological. This
may also
be seen from the fact that the expressions (2.1) and (2.5) for
these
phase shifts are independent of the shape of the curve
traveled by
the particle.
But notice that the argument above which establishes the
equality of
phase shifts in the two situations that are dual to each other
does {\it
not} assume that the phase shift is topological. It is valid for
the
phase shift due to any interaction, which may or may not be
topological. Also, the above duality can be generalized to the
case of the charge moving around a closed curve $C$ and
acquiring a
phase in the
field of an arbitrary distribution of dipoles, each having the
same
magnetic moment in both direction and magnitude. A little
thought,
by considering each pairwise interaction of the charge and
each
dipole, shows that in the dual situation, in which the dipole
moves
along the
parity transformed curve with the charges in the parity
transformed
positions of the dipoles of the original situation, the same
phase is
acquired
by the dipole. Again using the
invariance of this interaction under parity, it is concluded that
the
same phase shift is obtained when the dipole travels the
original
closed curve $C$ with the charges in the positions of the
dipoles in the original situation [9]. This argument may be
generalized to an arbitrary linear interaction, but the
interaction
needs to
be invariant under parity for the last step to be valid. The
equality
between
the two phases in the two dual situations will be called the
duality
principle.

This duality principle enables us to obtain from the known
phase shift due to a line source a new phase shift.
Alternatively, if
both phase shifts are known then this principle may be used
heuristically to obtain a new solution of the field equations
from the
old solution that gave the old phase shift. For example,
suppose we
know the
magnetic field of a solenoid and the AB phase shift [1] of a
charge
due to it,
and the phase shift of a magnetic moment due to a general
electric
field [2]. Then according to the duality principle, the phase
shift in
the
situation dual to the AB effect in which a charge is interfering
in the
electric
field ${\bf E}$ due to a line charge is
the same and is given by
(2.5). Then using the phase shift of the magnetic
moment due
to ${\bf E}$ implied by (1.4) and the axial symmetry one
obtains
${\bf
E}={\lambda\over
2\pi\rho}{\bf\hat\rho}$, where $\rho$ is the distance from
the line
charge, ${\bf\hat\rho}$ is a unit vector in the radial direction
and
$\lambda$ is the charge per unit length. Thus $\bf E$ of a line
charge
is obtained
without solving Maxwell's equations. In the next section, I
shall
apply this general argument to phase shifts produced by
gravity.

\vglue 0.6cm
\line{\elevenbf 3. Topological Phase Shifts in the Gravitational
Field \hfil}
\vglue 0.4cm

It is well known that the mass and the spin angular
momentum in a
gravitational field are, respectively, analogous to the charge
and
magnetic moment in an electromagnetic field. Therefore the
analog of
the AB effect for gravity is the phase shift $\Delta\phi_G$
acquired
by a mass going around a string that has angular momentum
(analog
of the solenoid). The dual situation then is a spin going around
a
string or a rod having mass only, and acquiring a phase
$\Delta\phi
'_G$.
Then according to the general argument in section 1, if the
field
equations
are linear,
$$\Delta\phi_G'=\Delta\phi_G . \eqno(3.1)$$

The actual values of $\Delta\phi_G$ and $\Delta\phi_G'$
depends on
the gravitational theory used to compute them. I shall study
these phase shifts in the following theories: A. Newtonian
gravity, B.
linearized limit of Einstein's theory of general relativity, and
C. The
Einstein-Cartan-Sciama-Kibble (ECSK) theory of the
gravitational field
with torsion [14]. In all three cases (3.1) will be shown to be
satisfied. The
differences between these phase shifts provide a way of
distinguishing, in
principle, between these
theories, although in practice the predicted effects are too
small for
realistic experimental tests at the present time.
\bigskip
\line{\elevenit A. Newtonian Gravity \hfil}
\smallskip
In this case, only the mass, not the angular momentum, acts
as the
source of gravity and is acted upon by gravity.  Therefore,
both
$\Delta\phi_G$ and $\Delta\phi_G'$ are zero. Hence, (3.1) is
trivially
satisfied.
\bigskip
\line{\elevenit B. Linearized General Relativity \hfil}
\smallskip
Consider now the low energy weak field limit of general
relativity.
Write the metric as
$g_{\mu\nu}=\eta_{\mu\nu}+\gamma_{\mu\nu}$,
where $\gamma_{\mu\nu}<<1$. In this subsection, all terms
which
are second order in $\gamma_{\mu\nu}$ will be neglected.
On
writing
$\overline\gamma_{\mu\nu}=\gamma_{\mu\nu}-
{1\over2}\eta_{\mu\nu}\eta^{\alpha\beta}\gamma_{\alpha\
beta}$,
the well
known linearized Einstein field equations are
$$\partial^\alpha\partial_\alpha\overline\gamma_{\mu\nu}=
8\pi
GT_{\mu\nu},\eqno(3.2)$$
in the gauge defined by
$\partial^\nu\overline\gamma_{\mu\nu}=0.$
I neglect stresses so that we have
$$T_{ij}=0, \overline\gamma_{ij}=0, i,j=1,2,3. \eqno(3.3)$$

Consider now a particle with mass $m$ and intrinsic spin
${\bf S}$ at
low
energies. In the stationary situation, a coordinate system may
be
chosen so
that $\gamma_{\mu\nu}$ are time independent. Then the
acceleration due
to gravity is
${\bf g}=-{1\over 2}\nabla \gamma_{00}$ and the `Coriolis'
vector
potential is
${\bf \gamma_0} =-
(\gamma_{01},\gamma_{02},\gamma_{03})$, as
can be
seen easily from the geodesic equation. The `gravi-
magnetic field' ${\bf
H}=\nabla\times {\bf \gamma_0} =2{\bf \Omega}$, where
${\bf
\Omega}$ is
the angular velocity of the coordinate basis relative to the
local
inertial frame. To couple the spin to the gravitational field
introduce
the
vierbein ${e^\mu}_a$ and its inverse ${e_\mu}^a$:
$${e^\mu}_a ={\delta^\mu}_a-{1\over 2}{\gamma^\mu}_a,
{e_\mu}^a
={\delta_\mu}^a+{1\over 2}{\gamma_\mu}^a ,\eqno(3.4)$$
which satisfy (3.20) and (3.22) below.
The latin indices, which take values $0,1,2,3$ may now be
lowered
and
raised using the Minkowski metric $\eta_{ab}$ and its inverse
$\eta^{ab}$.
It then follows that the Ricci rotation coefficients
${{\omega_\mu}^a}_b
\equiv {e_\nu}^a\nabla_\mu {e^\nu}_b$ are given by
$$\omega_{\mu ab}={1\over2}(\gamma_{\mu b,a}-
\gamma_{\mu
a,b}),\eqno(3.5)$$
where $,a$ denotes partial differentiation with respect to
$x^a$.

The phase shift in interference due to the gravitational field
may be
obtained
in the present approximation by taking the low energy weak
field
limit of
the phase shift obtained in reference 5. In particular, the
phase shift
due to
spin alone is obtained by parallel transporting the spin wave
function by
acting on it by the operator
$$\Phi_S =P\exp\left[- {i\over\hbar}\int_C
{1\over2}{{{\omega}_\mu}
^a}_b {S^b}_a dx^\mu
\right] =P\exp\left[- {i\over\hbar}\int_C
{1\over2}\gamma_{\mu
a,b} S^{ab}
dx^\mu
\right],\eqno(3.6)$$
where the integral is along the unperturbed classical
trajectory with
$P$
denoting path ordering, and ${S^b}_a$, which generate Lorentz
transformations in spin space, are related to the spin vector
$S^a$
and the 4-
velocity $v^b$ by
$$S^{ab} = \epsilon^{abcd}v_c S_d ,\eqno(3.7)$$
with all components being with respect to the vierbein. The
subsidiary condition $S^a v_a =0$ is assumed here.

Rewrite (3.6) as
$$\Phi_S =P\exp \left[-{i\over\hbar}\oint_C {1\over2}
\gamma_{0
a,b} S^{ab} dt
-{i\over\hbar}\oint_C
{1\over2}\gamma_{ia,b} S^{ab} dx^i \right] . \eqno(3.6')$$
The first integral in the exponent of $(3.6')$ is
$$\oint_C {1\over2}
\gamma_{00,i} S^{0i}dt +\oint_C {1\over 2}\gamma_{0i,j} S^{ij}
dt =
\oint_C ({\bf g}\times{\bf S}\cdot{\bf v}-{\bf\Omega}\cdot
{\bf S})dt . \eqno(3.8)$$
The second integral of $(3.6')$ is approximately
$$\oint_C {1\over2}\gamma_{ij,k} S^{jk} dx^i =\oint_C  {\bf
g}\times
{\bf S}\cdot d{\bf r}. \eqno(3.9)$$
Combining these results, (3.6) reads
$$\Phi_S = P\exp\left[-{i\over\hbar}2\oint_C  {\bf
g}\times{\bf
S}\cdot
d{\bf r}+
{i\over\hbar}\oint_C
{\bf\Omega}\cdot{\bf S}dt\right] .\eqno(3.10)$$
The precession represented by the last term of the exponent
in
(3.10)
corresponds to the interaction energy $-{\bf\Omega}\cdot{\bf
S}=-{1\over
2}{\bf S}\cdot{\bf H}$ in the Hamiltonian. This may be
understood
from the fact that when we transform to a frame rotating
with an
angular
velocity $\bf \Omega$ relative to the local inertial frame, the
spin
that is
constant in the inertial frame obviously
rotates with angular
velocity $-\bf \Omega$ relative to the new frame [15]. The
ratio
$\gamma$ of the magnetic moment to spin introduced in
section 1
for the electromagnetic interaction is, for a particle with
charge $e$
and mass $m$, $\gamma ={ge\over 2m}$, where $g$ is the
gyromagnetic ratio. For the gravitational field, the principle of
equivalence implies that the `charge' density equals mass
density.
Therefore,  $g=1$ and $e=m$. Hence, $\gamma ={1\over 2}$
and the
`gravi-
magnetic moment' ${\bf\mu}_G
= {1\over 2}{\bf S}$, consistent with the above interaction
energy.

Equations  (3.2)
subject to (3.3) are
$$\partial^\alpha\partial_\alpha\overline\gamma_{0\mu}=8\
pi
GT_{0\mu}.$$
These are like Maxwell's equations
in the Lorentz gauge,
and may be solved in the same way. Consider the specific case
of an
infinite
uniform hollow cylinder of radius $\rho_0$ and mass per unit
length
$\mu$ rotating
about its axis with angular momentum per
unit length $J$ parallel to the axis of the cylinder along the z-
axis.
This is analogous to a
rotating charged cylinder in electromagnetism. So, on defining
${\bf
r}=(x^1,x^2,x^3)$, ${\bf\rho} = (x^1,x^2,0)$, and
$\rho=|{\bf\rho}|$, the
solution exterior to the cylinder ($\rho >\rho_0$) is obtained
to be
$$\gamma_{00}=\gamma_{11}=\gamma_{22}=\gamma_{33}
=4G\mu \log\rho ,{\bf \gamma_0}=-{4G\over \rho^2}{\bf
J}\times
{\bf r}=-{4G\over \rho}{\bf J}\times {\bf \hat\rho}
,\eqno(3.11)$$
where $\bf\hat\rho$ is a unit vector in the direction of
$\bf\rho$.
The solution in the interior to the cylinder is
$$\gamma_{00}=\gamma_{11}=\gamma_{22}=\gamma_{33}
=4G\mu \log\rho_0 ,{\bf \gamma_0}=-{4G\over
\rho_0^2}{\bf
J}\times {\bf r}=-{4G\over \rho_0^2}{\bf J}\times {\bf \rho}
.\eqno(3.12)$$

Suppose at first that $J=0$. Then, from (3.11), ${\bf
\Omega}={\bf 0}$
 and ${\bf g}=-{2G\mu\over\rho}{\bf\hat\rho}$. Consider the
interference around the cylinder of a particle
whose spin is
polarized in
the $x^3-$direction with the axis of the cylinder lying along
the
$x^3$-
axis. Then
the
phase shift due to the coupling of spin to curvature [5,16] is
obtained
from
(3.10) to be
$$\Delta\phi_G=-{2\over\hbar}\oint_C  {\bf g}\times{\bf
S}\cdot
d{\bf r}=-
8\pi
{G\over
\hbar}\mu
S .\eqno(3.13)$$
This phase shift is independent of $C$ and is therefore
topological.

Consider now the dual situation that is constructed as follows.
Divide
the
cylinder into small segments of length $\delta
\ell$. The mass of each segment is $m=\mu\delta \ell$. In
performing the
duality operation, each segment is replaced by a segment
whose
spin is the same as $S$ and the particle is replaced with
another
particle with mass $m$. Then the cylinder has angular
momentum
per unit
length $J = {S\over\delta\ell}$. Therefore,
$${m\over\mu}={S\over J} .\eqno(3.14)$$
So, in the dual situation, the mass $m$ is interfering around
the cylinder with angular momentum per unit length ${\bf
J}$, which
is a gravitational analog of the AB
experiment. From (3.11), the phase shift due to ${\bf J}$ is
the
Sagnac phase
shift [4]
$$\Delta\phi '_G ={m\over\hbar}\oint_C{\bf\gamma_0}\cdot
d{\bf
r}=-8\pi
{G\over \hbar}mJ . \eqno(3.15)$$
It follows from (3.13), (3.14) and (3.15) that (3.1) is satisfied.

I shall now describe the gravitational analog
to the
topological phase shift of the neutron due to the
magnetic field
described towards the end of section 1.
Suppose, in a neutron or atomic
interference
experiment each of the two interfering
 beams passes along the axis of each of two identical very
massive
cylinders.
One of the cylinders rotates as the
wave
packet of each neutron enters the cylinder and
stops
rotating before the neutron leaves the cylinder.
Then from (3.12),
$${\bf \Omega}={1\over 2}\nabla\times {\bf \gamma_0}
=-{4G\over{\rho_0}^2}
{\bf J}.\eqno (3.16)$$
Suppose, for simplicity, that the spin of the neutron or atom is
polarized
along the axis of the cylinder. The first integral in (3.10)
would be
the same
for both beams. Therefore, the phase shift between them is
given by
the
second integral of (3.10) due to the rotating cylinder to be
$$\Delta\phi_\Omega = {1\over\hbar}\oint_C
{\bf\Omega}\cdot{\bf S}dt=
-{4GJS\tau\over\hbar{\rho_0}^2},\eqno(3.17)$$
on using (3.16), where $\tau$ is the time spent by the
neutron inside
the hollow
cylinder,
assuming that the time intervals during which the rotation of
the
cylinder is
turned on and off is negligible compared to $\tau$. As in the
electromagnetic case, the phase shift (3.17) is not
accompanied by a
force, apart from transient effects when the field is turned on
and off
which occurs also for the AB effect due to the scalar potential
[1].
Hence, (3.17) is a topological phase shift.

It is interesting that $\Delta\Phi_S$ obtained here by parallel
transport with respect to the gravitational connection is
analogous to
how the electromagnetic phase shift experienced by the
dipole was
obtained by parallel transport with respect to a corresponding
connection [8,9]. The three gravitational phase shifts obtained
above
using
(3.10) and (3.15) are the low energy weak field limit of the
phase
shifts obtained previously using Dirac's equation [5]. However,
these
phase factors correspond to the tentative Hamiltonian
$$H = {1\over 2m}({\bf p}-m{\bf \gamma}_0-2{\bf S}\times
{\bf
g})^2 + m{\gamma_{00}\over 2}-{1\over 2}{\bf S} \cdot {\bf
H} ,
\eqno(3.18)$$
in the sense that they may be derived from this Hamiltonian.
This is a generalization of the Hamiltonian found by DeWitt
[17] to
include spin. One way of confirming (3.18) is to directly take
the low
energy
weak field
limit of Dirac's equation, which will be studied in a future
paper.
\bigskip
\line{\elevenit C. General Relativity with Torsion \hfil}
\smallskip

The phase shift in general relativity may be obtained from
the action
on the wave function of the gravitational phase factor [18]
$$\Phi_C =  P\exp[- i\int_C ({e_\mu}^a
P_a +{1\over2}{{{\omega}_\mu} ^a}_b {M^b}_a)dx^\mu
],\eqno(3.19)$$
where
$C$ goes through the interfering beams.
Here,
$P_a$ and
${M^b}_a, a,b = 0,1,2,3$ are the translation and Lorentz
transformation generators which generate the Poincare group
that
acts on the Hilbert space. Then ${e_\mu}^a$ and
${{{\omega}_\mu}^a}_b$ have the interpretation of the gauge
potentials of a
Poincare gauge field. In an interferometry experiment the
two beams
need to be brought together by means of mirrors which gives
rise to
the Thomas precession [19], which will be treated elsewhere
[20].

It was shown by means of the WKB approximation of Dirac's
equation that (3.19) determines the gravitational phase [5,18].
This
may also
be realized for a particle with arbitrary spin as follows:  The
Lorentz
part of
(3.19) ensures that the wave packet is parallel
transported infinitesimally, while it acquires a phase, which is
a good
approximation for the locally
approximate plane wave being considered here. To find the
phase
acquired
due to energy-momentum, note first that ${e_\mu}^a$
depends on
the
observer. A Lorentz
transformation of the observer results in ${e_\mu}^a$
transforming as a contravariant vector in the index $a$ while
$P_a$ transforms as a covariant vector. Suppose that a
particle is
in a state $|\psi>$ that is approximately an eigenstate of $P_a$
with eigenvalues $p_a$. The fact that the gravitational phase
is
observable along an open curve implies
that the ``wave vector'' $p_\mu \equiv {e_\mu}^a p_a$ is
observable [21,22]. Requiring that the correspondence
between
$p_\mu$
and $p_a$ is $(1-1)$ implies that ${e_\mu}^a$ is a non
singular
matrix. Therefore, it has an inverse ${e^\mu}_b$:
$${e_\mu}^a{e^\mu}_b ={\delta^a}_b . \eqno(3.20)$$
Hence, $p_a = {e^\mu}_a p_\mu$. The Casimir
operator $\eta^{ab}P_aP_b$ of the
Poincare group has a definite value, say $m^2$, for the given
particle. Therefore,
$$\eta^{ab}p_ap_b= g^{\mu\nu}p_\mu p_\nu = m^2 ,
\eqno(3.21)$$
where $g^{\mu\nu}\equiv \eta^{ab}{e^\mu}_a {e ^\nu}_b$ is a
non
singular matrix. Its inverse
$g_{\mu\nu}$ defines a pseudo-Riemannian metric of
Lorentzian
signature on space-time. On using (3.20),
$$g_{\mu\nu}=\eta_{ab}{e_\mu}^a {e_\nu}^b .\eqno(3.22)$$
Thus the definiteness of the mass (which may be zero)
ensures the
definiteness of the phase that depends on ${e_\mu}^a$ even
for open
curves. In this way the space-time metric is deduced from the
gravitational phase corresponding to the translational part of
(3.19)
which is observable along an open curve. Conversely, the
metric
determines the latter phase to be observable along an open
curve.

The field strength or curvature of this Poincare gauge field is
obtained by evaluating (3.19) for an infinitesimal closed
curve
$C$:
$$\Phi_C  =  1-{i\over 2} ({Q_{\mu \nu}} ^a
P_a +{1\over 2}{{R_{\mu \nu}}^a}_b {M^b}_a)d\sigma ^{\mu
\nu},\eqno(3.23)$$
using the Poincare Lie algebra, where
$$Q^a=de^a+{\omega
^a}_b\wedge e^b \eqno(3.24)$$
is the torsion and
$${R^a}_b={d\omega ^a}_b+{\omega ^a}_c\wedge {\omega
^c}_b
\eqno(3.25)$$ is the linear
curvature.

Comparison of (3.19) with (1.2) and (1.4), and (3.23) suggest
that
there
may be topological phase shifts due to interference of
coherent
beams that enclose a region that contains curvature and
torsion, but
which are zero along the beams. Such an example is provided
by the
cosmic string
whose metric exterior to the string is
given cylindrical coordinates as [23, 24, 25].
$$ds^2 = (dt + \beta\,d\phi )^2 -
d\rho^2-\alpha^2 \rho^2 d\phi^2-dz^2 ,\eqno(3.26)$$
where $\alpha$ and $\beta$ constants. Then
the metric $g_{\mu\nu}$ satisfies (3.22) for the following
orthonormal co-frame field $\{e ^a\}$ adapted to the above
coordinate system:
$$e^0= dt+\beta d\phi , e^1 =d\rho , e^2 =\alpha\rho d\phi ,
e^3=dz
.\eqno(3.27)$$
The connection coefficients in this basis are
${{\omega_\mu}^a}_b \equiv {e_\nu}^a\nabla_\mu
{e^\nu}_b=0$,
for all $a,b,\mu$ except for
$${{\omega_{\hat \phi}}^1}_2=-{{\omega_{\hat \phi}}^2}_1 =-
\alpha d\phi. \eqno(3.28))$$
It follows, on using (3.24) and (3.25), that $Q^a=0,{R^a}_b=0$
outside
the string. The scattering cross section of particles with
definite
energy in the above geometry has been obtained before [26].

In the appendix it is shown that this solution may be
extended to an
interior solution that has uniform energy and spin densities
and
which generate curvature and torsion according to the ECSK
equations [14]. The constants $\alpha$ and $\beta$ are then
determined
by matching the interior and exterior solutions to be
$$\alpha =1-4G\mu, \beta=4GJ, \eqno(3.29)$$
where $\mu$ is the mass per unit length and $J$ is the
angular
momentum per unit length due to the intrinsic spin density
inside
the string.

If $C$ is a closed curve around the cosmic string then from
(3.19),
$$\Phi_C =  \exp\left(- i\oint_{C} {e_\mu}^0
P_0 dx^\mu \right) P\exp\left[- i\oint_{C}
\left(\sum_{k=1}^3{e_\mu}^k
P_k+{{{\omega}_\mu} ^1}_2 {M^2}_1\right)dx^\mu
\right],\eqno(3.30)$$
using the fact that ${e_\mu}^0P_0$ commutes with
the other
terms that occur in (3.30).The first exponential is a time
translation second is a spatial Euclidean transformation.
Hence, if
(3.23) is valid, then the time
translation would correspond to torsion being non zero inside
the
string. Suppose that the surface of the string is given by
$\rho=\rho_0$, where $\rho_0$ is a small constant. Then,
substituting (3.27),
(3.28) into (3.30), and using (3.23), the flux of torsion through
a
cross-section
$\Sigma$ of the string is
$$\int_\Sigma Q^0=2\pi\beta ,\int_\Sigma {R^1}_2=2\pi (1-
\alpha)
\eqno(3.31)$$
This is independent of the particular geometry inside the
string
so long as $\Sigma$ is
``infinitesimal'' so that (3.23) is valid. In particular, (3.31) is
easily
verified for
the solution in the appendix, independently of the value of
$\rho_-$,
using
(A.12), (A.15) and (A.16).

For simplicity, consider a circular interferometer with
constant
radius $r>\rho_0$ in a plane normal to the string, with its
center
on the axis of the string. It may be a superconducting
interferometer, e. g. a superconducting ring interrupted by a
Josephson junction. Or it may be an electron interferometer,
or a
wave guide,
such as an optical fiber, at
one point of which is the beam splitter that splits a beam into
two
which travel in opposite senses and interfere at a mirror that
is at
another point in the interferometer. The interferometer does
not
rotate relative to the distant stars, which may be ensured by
requiring that telescopes rigidly attached to this
interferometer
are focused on the distant stars.

The phase shift may be obtained using (3.30) with $C$ along
integral curves of $p^\mu$ which lie on a 2 dimensional
submanifold $\sigma$ with constant $z$ and $\rho=r$. Hence,
$C$
may be chosen to be along a circle around $\sigma$ with
constant
$t$. Suppose $E$ is the energy of the wave function which
is
assumed to be constant in time at the beam splitter. Then it is
constant everywhere along the beams. Therefore, in this WKB
approximation, the
magnitude
of the momentum $p=(E^2-m^2)^{1/2}$ is also a
constant along the beam. By taking into account the Fermi-
Walker
transport of vectors associated with $|\psi>$, ${M^2}_1$ in
(3.30) may
be replaced by the spin operator ${S^2}_1$ in the present
coordinate
basis [19]. The spin is
assumed to be polarized in the $z$-direction, i. e. $|\psi>$ is an
eigenvector of ${S^2}_1$ with eigenvalue $S/\hbar$.

Now, the three
operators in (3.30) commute with one another and their
actions on
$|\psi>$ give rise to the following topological phase shifts:
(i) The general relativistic Sagnac phase shift [4] is
obtained from the first factor to be
$$\Delta \phi_E=-\oint_{C} {E\over\hbar}{e_{\hat\phi}}^0
 d\phi=-\int_\Sigma {E\over\hbar}{Q_{\hat\rho \hat\phi}}^0
d\rho\wedge d\phi=-8\pi {G\over\hbar} EJ ,
\eqno(3.32)$$
where $\Sigma$ is a 2-surface spanned by $C$.
(ii) The phase shift due to the coupling of spin to curvature [5]
which is obtained by the second factor in
(3.30)\footnote{*}{\ninerm\baselineskip=11pt
The phase
shifts (3.32) and (3.33) may be evaluated using the line
integral
outside the string using (A.16) or
the surface integral inside the string using (A.11). In (3.33),
$2\pi$
has been subtracted from the
line integral to remove the purely coordinate effect due to the
rotation of $e^2$ by $2\pi$ as one goes around $C$, consistent
with
the Gauss-Bonnet theorem.\hfil}
$$\Delta\phi_S =-{S\over\hbar}\{\oint_C
{{{\omega}_{\hat\phi}} ^1}_2  d\phi -2\pi\}=-\int_\Sigma
{{R_{\hat\rho \hat\phi}} ^1}_2 d\rho\wedge
d\phi=-8\pi {G\over\hbar}S\mu . \eqno(3.33)$$

The phase shifts (3.32) and (3.33) are expressed as flux
integrals
of torsion and curvature because they could have also been
obtained from (3.19) which depends only on the affine
connection.
(The torsion and curvature fluxes contained in (3.32) and
(3.33) are
the same as (3.31) which is
independent of the particular geometry interior to the string
onsidered here.) It
follows that these phase shifts are independent of the shape
of the
interferometer enclosing the string and therefore may be
called
topological.

I shall now show that the above topological effects satisfy the
principle of duality formulated in section 2. Consider first the
Sagnac
effect on a particle with energy $E$ due to the spinning string
with
angular momentum per unit length $J$. This is like the AB
effect due
to a solenoid. Divide the string into small segments of length
$\delta
\ell$. The spin of each segment is $S=J\delta \ell$. In
performing the
duality operation, each segment is replaced by a segment
whose
mass is the same as $E$ and the particle is replaced with
another
particle with spin $S$. Then the solenoid has been replaced by
a rod
with mass per unit length $\mu = {E\over\delta\ell}$.
Therefore,
$${E\over\mu}={S\over J} .\eqno(3.34)$$
Conversely, if (3.34) is valid then the two situations may be
obtained
from each other by performing the duality operation. Hence,
by the
duality principle, the phase shifts for the two situations
should be
equal. Indeed, the phase shifts (3.32) and (3.33) which were
derived
without paying any attention to the duality principle are
equal if and only if (3.34) is valid.

This illustrates also again how the duality principle may be
used to
obtain
the phase shift for the dual situation: From (3.32), we may
obtain
(3.33), or vice versa, on using (3.34). Even though the general
relativistic equations are in general non linear, the equations
that are
solved in the appendix to obtain the exact solution are all
linear, so
that there must be duality in the present case according to
the
general arguments of section 2. If this duality is assumed
then a new
gravitational solution may be obtained from an old solution
both in
the present case and in the low energy weak field case
considered
earlier, similar to how this was done in the electromagnetic
case at
the end of section 2.

\vglue 0.6cm
\line{\elevenbf 4. Concluding Remarks \hfil}
\vglue 0.4cm

As already mentioned, the AB effect shows that the field
strength is
insufficient to describe the electromagnetic field, whereas the
phase
factor (1.2), which is called a holonomy transformation
because it
parallel transports around a closed curve, adequately
describes the field. More generally, for an arbitrary gauge
field, the
holonomy transformations are of the form (1.4), with
${A_\mu}^k$ now being the corresponding vector potential.
The fact
that these are
sufficient to determine the field uniquely is
shown by the following theorem [27]: Given the holonomy
transformations (2) for piece-wise differentiable curves
which
begin and end at a given point in space-time, the gauge
potential
${A_\mu}^k$ may be reconstructed, and it is then unique up
to
gauge transformations.

But since the set of such curves form an
infinite dimensional manifold {\cal L}, the corresponding
operators (2) have a great deal of redundancy. Indeed, the
gauge
field in space-time may be reconstructed
from a minimal set of these operators defined on a four
dimensional submanifold
of {\cal L} [7]. This is mathematically
equivalent to working with the gauge potential defined in a
particular gauge on the four dimensional space-time.
Therefore,
once the redundancy in the loop space {\cal L} has been
removed, there is
no advantage to using the holonomy transformation (2) as
opposed to the gauge potential in a particular gauge.

It follows that in quantizing the
electromagnetic or more general gauge fields, one must
quantize
the gauge potential instead of the field strength. Similarly, the
topological effects due to the gravitational field described
above
suggest that in quantizing the gravitational field, it is the
`gauge
potentials' ${e_\mu}^a$ and ${M^b}_a$ which should be
quantized,
and the metric (3.22) is obtained from them as a secondary
variable
[26,22]. However, there is a breaking of gauge symmetry
which
makes ${e_\mu}^a$ a tensor field instead of a connection [22].
This is
like how in a superconductor the $U(1)$ gauge symmetry is
spontaneously broken, which makes $A_\mu$ a covariant
vector
field instead of a connection.

So, it may well be that in the early universe there was the full
Poincare gauge symmetry with ${e_\mu}^a$ and ${M^b}_a$
having
vacuum expectation value zero in an appropriate gauge. As a
result
of spontaneous symmetry breaking of the translational part
of the
Poincare group, ${e_\mu}^a$ may have acquired a vacuum
expectation value equal to ${\delta_\mu}^a$ corresponding to
the
Minkowski geometry. But I emphasize that these are
speculative
remarks, and need justification by a detailed theory.
\vglue 0.6cm
\line{\elevenbf Acknowledgements \hfil}

I thank Y. Aharonov and P. O. Mazur for stimulating
discussions
and J. L. Safko for several enlightening discussions. This work
was
partially
supported by
NSF grant PHY-9307708 and ONR grant R\&T 3124141.
\vglue 0.8cm
\line{\elevenbf Appendix: Spinning Torsion String \hfil}
\bigskip

The simplest gravitational field
equations in the presence of torsion are the Einstein-Cartan-
Sciama-
Kibble (ECSK) equations [14], which may be written in the
form
[29]
$${1\over 2}\eta_{ijkl}\theta^l\wedge R^{jk}=-8\pi G t_i
,\eqno(A.1)$$
$$\eta_{ijkl}\theta^l\wedge Q^k = 8\pi G s_{ij} ,\eqno(A.2)$$
where $t_i$ and $s_{ij}$ are 3-form fields representing the
energy-momentum and spin densities. I shall now obtain an
exact
solution of these equations for the interior of the cosmic
string
which
matches the exterior solution (3.26). This will then give
physical
and geometrical meaning to the parameters $\alpha$ and
$\beta$
in (3.26). This
solution will be different from earlier torsion string solutions
[31]
that have static interior metrics matched with exterior
metrics which are different from (3.26).

The $\rho$ and $z$ coordinates in the interior will be chosen
to be
the distances measured by the metric in these directions.
Since the
exterior solution has symmetries in the $t, \phi$, and $z$
directions,
it is reasonable to suppose the same for the interior solution.
So,
all functions in the interior will be functions of
$\rho$ only. Requiring also simplicity, I make the following
ansatz
in
the interior:
$$\theta^0 =u(\rho) dt + v(\rho) d\phi , \theta^1 = d\rho ,
\theta^2
=
f(\rho) d\phi , \theta^3 = dz , {{\omega} ^2}_1=k(\rho) d\phi =
-{{\omega} ^1}_2 ,\eqno(A.3)$$
all other components of ${{\omega} ^a}_b$ being zero, and
$ds^2=\eta_{ab}\theta^a\theta^b \equiv g_{\mu\nu}dx^\mu
dx^\nu$. Suppose also that the energy density $\epsilon$ and
spin
density $\sigma$, polarized in the $z$-direction, are constant
and
classical fluid at rest. I. e.
$$\eqalignno{t_0 &= \epsilon \theta^1\wedge \theta^2
\wedge
\theta^3 = \epsilon
f(\rho)d\rho\wedge d\phi \wedge dz ,\cr
s_{12} =-s_{21}&=\sigma
\theta^1\wedge \theta^2 \wedge \theta^3 = \sigma
f(\rho)d\rho\wedge d\phi \wedge dz ,  &(A.4)\cr}$$
the other components of $s_{ij}$ being zero.
In terms of the components of the energy-momentum and
spin
tensors in the present basis, this means that ${t^0}_0
=\epsilon=$
constant and ${s^0}_{12} = \sigma =$ constant.

It is assumed that there is no surface energy-momentum or
spin
for
the string. Then the metric must satisfy the junction
conditions
[30], which in the present case are
$$g_{\mu\nu}|_-=g_{\mu\nu}|_+ , \, \partial_{\hat\rho}
g_{\mu\nu}|_+
=\partial_{\hat\rho} g_{\mu\nu}|_- + 2K_{(\mu\nu)\hat\rho}
,
\eqno(A.5) $$
where
$K_{\alpha\beta\gamma} = {1\over 2}(-
Q_{\alpha\beta\gamma}+Q_{\beta\gamma\alpha}-
Q_{\gamma\alpha\beta})$ is the contorsion or the defect
tensor,
$|_+$ and $|_-$ refer to the limiting values as the boundary of
the
string is approached from outside and inside the string,
respectively,
and the hat denotes the corresponding coordinate component.

Substitute (A.3), (A.4) into the Cartan equations (A.2). The
$(i,j)=(0,2),$
$(0,3), (2,3)$ eqs. are automatically satisfied.
The $(i,j)=(0,1),(1,3),(1,2)$ eqs. yield
$$f'(\rho)-k(\rho) = 0, u'(\rho) = 0 , v'(\rho) = 8\pi G\sigma
f(\rho), \eqno(A.6)$$
where the prime denotes differentiation with respect to
$\rho$.
Therefore, the continuity of the metric (eq. (A.5)) implies that,
since $u=1$ at the boundary, $u(\rho)=1$ everywhere. Now
substitute (A.3), (A.4) into the Einstein equations (A.1). The
$i=0$
eq. yields
$$k'(\rho) = -8\pi G\epsilon f(\rho) . \eqno(A.7)$$
The $i=1,2,3$ equations yield, respectively
$$t_1=0,t_2=0,t_3
= {k'\over 8\pi G}dt\wedge d\rho\wedge d\phi
=-
\epsilon\theta^0\wedge\theta^1\wedge\theta^2,\eqno(A.8)$$
using (A.7). Hence, ${t^3}_3 = \epsilon = {t^0}_0$. From (A.6)
and
(A.7),
$$f''(\rho) + {1\over {\rho *}^2}f(\rho) = 0 , \eqno(A.9) $$
where $\rho * =(8\pi G\epsilon )^{-1/2}$.
In order for there not to be a metrical ``cone'' singularity at
$\rho
= 0$, it is necessary that
$\theta^2\sim \rho d\phi$ near $\rho =0$.
Hence, the solution of (A.9) is $f(\rho)
= \rho* sin{\rho\over \rho *}. $
Then from (A.6), $k(\rho)=cos{\rho\over \rho *}$, and
requiring
$v(0)=0$ to avoid a conical
singularity, $v(\rho) = 8\pi G\sigma \rho *^2\left(1-
cos{\rho\over \rho *}\right). $
This gives the metric in the interior of the string to be
$$ds^2 = \left[dt + 8\pi G\sigma \rho *^2\left(1-
cos{\rho\over
\rho*}\right)\right]^2
-d\rho^2 -\rho*^2 sin^2\left({\rho\over \rho *}\right) d\phi^2
-dz^2.
\eqno(A.10)$$
The only non vanishing components of curvature and torsion
are
$$Q^0 = 8\pi G\sigma\rho *
sin\left({\rho\over \rho *}\right) d\rho\wedge
d\phi , {R^1}_2={1\over\rho*}sin\left({\rho\over \rho
*}\right)
d\rho\wedge
d\phi=-{R^2}_1.\eqno(A.11)$$

I apply now the junction conditions (A.5), which will
show that $\rho$ is discontinuous across the boundary.
Denote the
values of
$\rho$ for the boundary in the internal and external
coordinate
systems by $\rho_-$ and $\rho_+$ respectively.
{}From (3.22) and (A.10), $g_{\hat t\hat \phi}$ and $g_{\hat
\phi\hat
\phi}$
are respectively continuous iff
$$ \beta = 8\pi G\sigma \rho *^2
\left(1- cos{\rho_-\over \rho *}\right) ,
\eqno(A.12)$$
$$\alpha \rho_+ = \rho * sin{\rho_-\over\rho *}.
\eqno(A.13)$$
The remaining metric coefficients are clearly continuous. The
only
non zero contorsion terms which enter into (A.5) are
obtained from (A.11) to be
$$K_{(\hat\phi\hat t)\hat\rho}=-4\pi G\sigma\rho *
sin{\rho\over\rho *}, K_{\hat\phi\hat\phi\hat\rho}=-(8\pi
G\sigma)^2 \rho *^3\left(1- cos{\rho\over \rho *}\right)
sin{\rho\over\rho *}.\eqno(A.14)$$
Using (A.13) and(A.14), it can now be verified that the
remaining
junction conditions (A.5) are satisfied provided
$\alpha = cos{\rho_-\over\rho *}.$
The mass per unit length is
$$\mu \equiv \int_\Sigma \epsilon\theta^1\wedge\theta^2
={1\over
4G}\left(1-cos{\rho_-\over\rho *}\right) ={1\over 8\pi
G}\int_\Sigma {R^1}_2, \eqno(A.15)$$
where $\Sigma$ is a cross-section of the string (constant
$t,z$).
Therefore,
$\alpha = 1-4G\mu$. The angular momentum per unit length
due
to
the spin density is
$$J\equiv \int _\Sigma\sigma\theta^1\wedge\theta^2
=2\pi\sigma\rho *^2\left(1-cos{\rho_-\over\rho
*}\right)={1\over 8\pi G}\int_\Sigma Q^0.\eqno(A.16)$$
Hence, from (A.12), $\beta =4GJ$. The Sagnac phase shift
obtained
earlier is therefore $\Delta\phi = ET$, where $T$ is the flux of
$Q^0$ through $\Sigma$. In the special case when torsion is
absent, which in the ECSK theory
means that spin density is zero, $\beta =0$, and the above
solution
reduces to the exact static solution of Einstein's theory found
by
Gott
[32] and others [33], whose linearized limit was previously
found by Vilenkin [34]. After this work was completed I
learned that
Tod [35] has studied torsion singularities using affine
holonomy and
the ECSK equations
analogous to the present approach.

\noindent
{\it Note added in proofs}: The phase shift due to spin in the
interference around a rod and a cosmic string has also been
studied
by B. Reznik (PhD thesis, Tel Aviv Univ., 1994, and preprint to
be
published in Phys. Rev. D), by using the contribution to the
Lagrangian due to the gravitational
interaction energy $U = {1\over2}\int
T^{\mu\nu}\gamma_{\mu\nu} d^3x$. This amounts to
treating
gravity as a spin 2 field, compared to the present geometric
approach
which begins with the full general relativistic theory.
However, the
above mentioned paper assumes that $T^{oi}$ in the rest
frame of the
particle is
the curl of spin density, which is then boosted to the
laboratory
frame. This assumption corresponds to setting the `gravi-
magnetic moment' $\mu_G$ equal to the spin. This differs
from the
result in section 3 of the present paper that $\mu_G$ is half
the spin
in accordance with the principle of equivalence. The latter
result
implies that $T^{oi}$ in the rest frame is half the curl of spin
density.
Then, integrating by parts, it is easy to show that the
Lagrangian for
a particle with mass $m$, velocity $\bf v$ and spin $\bf S$
in the laboratory frame is
$$L\equiv {1\over 2}m{\bf v}^2-U = {1\over 2}m{\bf v}^2 -
m{\gamma_{00}\over 2}+m{\bf v}\cdot {\bf \gamma}_0+
2{\bf v}\cdot {\bf S}\times {\bf g}  + {1\over 2}{\bf S} \cdot
{\bf
H} .$$
This confirms the Hamiltonian (3.18) of the present paper.
Also, the present paper studies an additional spin interaction
represented by the last term of
(3.18), or the last term of the above Lagrangian, which gives
rise to a
new topological
phase
shift (3.17).

\vglue 0.6cm
\line{\elevenbf References \hfil}
\vglue 0.4cm

\item{1.} Y. Aharonov and D. Bohm, Phys. Rev. {\bf 115,} 485
(1959).
\item{2.} J. Anandan, Phys. Rev. Lett. {\bf  24,} 1660 (1982).
\item{3.} Y. Aharonov and A. Casher, Phys. Rev. Lett. {\bf 53,}
319
(1984).
\item{4.} J. Anandan, Phys. Rev. D {\bf 15,} 1448 (1977).
\item{5.} J. Anandan, Nuov. Cim. A {\bf 53,} 221
(1979).
\item{6.} T. T. Wu and C. N. Yang, Phys. Rev. D, {\bf 12,}  3845
(1975).
\item{7.} J. Anandan, Phys. Rev. D {\bf 33,} 2280 (1986).
\item{8.} J. Anandan, Phys. Lett. A {\bf 138,} 347 (1989).
\item{9.}J. Anandan in {\it Proceedings of the International
symposium
on the Foundations of Quantum Mechanics,} Tokyo, August
1989,
edited by S. Kobayashi et al (Physical Society of Japan, 1990)
P. 98.
\item{10.} A. S. Goldhaber, Phys. Rev. Lett. {\bf 62,} 380
(1989).
\item{11.} D. Wisnievsky and Y. Aharonov, Ann. Phys. {\bf
45,} 479
(1967).
\item{12.} A. Zeilinger in {\it Fundamental Aspects of
Quantum
Theroy,}
edited by V. Gorini and A. Frigerio (Plenum, NY 1985).
\item{13.} B. E. Allman et al, Phys. Rev. Lett. {\bf 68,} 2409
(1992);
errata, Phys.
Rev. Lett. {\bf 70,} 250 (1993).
\item{14.} D. W. S. Sciama in Recent Developments in General
Relativity (Oxford 1962). p. 415; T. W. B. Kibble, J. Math. Phys.
{\bf
2,} 212
(1961).
\item{15.} See, for example, J. Anandan, Phys. Rev. Lett. {\bf
68,}
3809 (1992).
\item{16.} J. Anandan and B.
Lesche, Lettre al Nuovo Cimento {\bf 37,} 391 (1983).
\item{17.} B. DeWitt, Phys. Rev. Lett. {\bf 16,} 1092 (1966).
\item{18.} J. Anandan in {\it Quantum Theory and
Gravitation},
edited by A. R. Marlow (Academic Press, New York 1980), p.
157.
\item{19.} J. Anandan, Phys. Rev. D {\bf 24,} 338 (1981) sect.
3.
\item{20.} J. Anandan, to be published in Phys. Lett. A (1994).
\item{21.} J. Anandan in
{\it Topological Properties and Global Structure of Space-
Time,}
 eds. P. G.
Bergmann and V. De Sabbata (Plenum Press, NY 1985), p. 1-
14; J.
Anandan, Ann. Inst. Henri Poincare {\bf 49,} 271 (1988).
\item{22.} J. Anandan in Directions in Directions in General
Relativity, Volume 1, Papers in honor of Charles Misner,
edited by
B. L. Hu, M. P. Ryan and C. V. Vishveshwara (Cambridge Univ.
Press, 1993).
\item{23.} L. Marder, Proc.
Roy. Soc. A {\bf 252,} 45 (1959) and in Recent
Devolopments in General Relativity (Pergamon, New York
1962); A.
Staruskiewicz, Acta. Phys. Pol. {\bf 424,} 734 (1963);
J. S. Dowker, Nuov. Cim. B {\bf 52,} 129 (1967);  J.
L. Safko and L. Witten, Phys. Rev. D {\bf 5,} 293 (1972); V. B.
Bezerra, J. Math. Phys. {\bf 30,} 2895 (1989) .
\item{24.}S. Deser, R. Jackiw, and G. 't Hooft, Ann. Phys. NY
{\bf 152,}
220
(1984);
\item{25.} P. O. Mazur, Phys. Rev. Lett. {\bf 57,} 929 (1986).
\item{26.} P. O. Mazur, Phs. Rev. Lett. {\bf 59,} 2380
(1987).
\item{27.} J. Anandan in {\it Conference on Differential
Geometric
Methods
in Physics}, edited by G. Denardo and H. D. Doebner (World
Scientific,
Singapore, 1983) p. 211.
\item{28.} J. Anandan, Found. Phys. {\bf 10,} 601 (1980).
\item{29.} A. Trautman in {\it The Physicist's Conception of
Nature},
edited
by J. Mehra (Reidel, Holland, 1973).
\item{30.} W. Arkuszewski, W. Kopczynski, and V. N.
Ponomariev,
Commun. Math. Phys. {\bf 45,} 183 (1975).
\item{31.}A. R. Prasanna, Phys. Rev. D {\bf 11,} 2083 (1975);
D.
Tsoubelis, Phys. Rev. Lett. {\bf 51,} 2235 (1983).
\item{32.} J. R. Gott III, Astrophys. J. {\bf 288,} 422 (1985).
\item{33.}W. A. Hiscock,
Phys. Rev. D {\bf 31,} 3288 (1985); B. Linet, Gen. Rel. and
Grav. {\bf
17,}
1109 (1985).
\item{34.} A. Vilenkin, Phys. Rev. D {\bf 23,} 852 (1981).
\item{35.} K. P. Tod, Class. and Quantum Grav. {\bf 11,} 1331
(1994).

\vfil\eject
\bye